\documentclass[sn-mathphys, Numbered, iicol]{sn-jnl_arx}
\usepackage{graphicx}
\usepackage{multirow}
\usepackage{amsmath,amssymb,amsfonts}
\usepackage{amsthm}
\usepackage{mathrsfs}
\usepackage[title]{appendix}
\usepackage{xcolor}
\usepackage{textcomp}
\usepackage{manyfoot}
\usepackage{booktabs}
\usepackage{algorithm}
\usepackage[figuresright]{rotating}
\usepackage{algorithmicx}
\usepackage{algpseudocode}
\usepackage{listings}
\usepackage{cleveref}
\usepackage{ulem}
\usepackage{xspace}
\usepackage{mhchem}
\usepackage{subcaption}
\usepackage{multirow}
\usepackage{array}
\usepackage[switch,displaymath, mathlines]{lineno}
\makeatletter
\newcommand*{\myfnsymbolsingle}[1]{
  \ensuremath{
    \ifcase#1
    \or
      \S
    \or
      \dagger
    \or
      a
    \or 
      b
    \or
      c
    \or
      d
    \or
      e
    \or
      f
    \or
      g
    \else
      \@ctrerr  
    \fi
  }   
}   
\makeatother

\usepackage{alphalph}
\newalphalph{\myfnsymbolmult}[mult]{\myfnsymbolsingle}{}

\usepackage{fnpct}
\theoremstyle{thmstylethree}

\Crefname{table}{\text{Table}}{\text{Tables}}
\Crefname{figure}{\text{Fig.}}{\text{Figs.}}
\Crefname{section}{\text{Section}}{\text{Sections}}
\Crefname{equation}{\text{Eq.}}{\text{Eqs.}}

\newcommand{\sptable}[1]{Supplementary Table #1}
\newcommand{\spfigure}[1]{Supplementary Figure #1}

\newcommand{\spsection}[1]{Supplementary Section #1}

\begin{document}

\title[Article Title]{Knowledge-driven AI-generated data for accurate and interpretable breast ultrasound diagnoses}

\author[1]{\fnm{Haojun} \sur{Yu}}\nomail\intern{Yizhun Medical AI}\equalcont{These authors contributed equally to this work.}
\author[1]{\fnm{Youcheng} \sur{Li}}\nomail\intern{Yizhun Medical AI}\equalcont{These authors contributed equally to this work.}
\author[2]{\fnm{Nan} \sur{Zhang}}\nomail\equalcont{These authors contributed equally to this work.}
\author[3]{\fnm{Zihan} \sur{Niu}}\nomail\equalcont{These authors contributed equally to this work.}
\author[4]{\fnm{Xuantong} \sur{Gong}}\nomail
\author[3]{\fnm{Yanwen} \sur{Luo}}\nomail
\author[1]{\fnm{Quanlin} \sur{Wu}}\nomail
\author[2]{\fnm{Wangyan} \sur{Qin}}\nomail
\author[3]{\fnm{Mengyuan} \sur{Zhou}}\nomail
\author[4]{\fnm{Jie} \sur{Han}}\nomail
\author[3]{\fnm{Jia} \sur{Tao}}\nomail
\author[6]{\fnm{Ziwei} \sur{Zhao}}\nomail
\author[1]{\fnm{Di} \sur{Dai}}\nomail
\author[1,a]{\fnm{Di} \sur{He}}\email{dihe@pku.edu.cn}
\author[6,b]{\fnm{Dong} \sur{Wang}}\email{dong.wang@yizhun-ai.com}
\author[5,c]{\fnm{Binghui} \sur{Tang}}\email{tbh691203@163.com}
\author[2,d]{\fnm{Ling} \sur{Huo}}\email{hlbcus@163.com}
\author[3,e]{\fnm{Qingli} \sur{Zhu}}\email{zhuqingli@pumch.cn}
\author[4,f]{\fnm{Yong} \sur{Wang}}\email{wangyong@cicams.ac.cn}
\author[1,g]{\fnm{Liwei} \sur{Wang}}\email{wanglw@pku.edu.cn}

\affil[1]{\orgname{Peking University}}
\affil[2]{\orgname{Peking University Cancer Hospital \& Institute}}
\affil[3]{\orgname{Peking Union Medical College Hospital}}
\affil[4]{\orgname{Cancer Hospital, Chinese Academy of Medical Sciences}}
\affil[5]{\orgname{Nanchang People's Hospital}}
\affil[6]{\orgname{Yizhun Medical AI Co., Ltd}}

\abstract{
    Data-driven deep learning models have shown great capabilities to assist radiologists in breast ultrasound (US) diagnoses. However, their effectiveness is limited by the long-tail distribution of training data, which leads to inaccuracies in rare cases. In this study, we address a long-standing challenge of improving the diagnostic model performance on rare cases using long-tailed data. Specifically, we introduce a pipeline, TAILOR, that builds a knowledge-driven generative model to produce tailored synthetic data. The generative model, using 3,749 lesions as source data, can generate millions of breast-US images, especially for error-prone rare cases. The generated data can be further used to build a diagnostic model for accurate and interpretable diagnoses. In the prospective external evaluation, our diagnostic model outperforms the average performance of nine radiologists by 33.5\% in specificity with the same sensitivity, improving their performance by providing predictions with an interpretable decision-making process. Moreover, on ductal carcinoma in situ (DCIS), our diagnostic model outperforms all radiologists by a large margin, with only 34 DCIS lesions in the source data. We believe that TAILOR can potentially be extended to various diseases and imaging modalities.
}

\maketitle
\small

\section{Main}
\label{sec:main}

\stepcounter{footnote}\footnotetext{These authors carried out this work as interns at Yizhun Medical AI Co., Ltd.}
\stepcounter{footnote}\footnotetext{Equal Contribution}
\stepcounter{footnote}\footnotetext{dihe@pku.edu.cn}
\stepcounter{footnote}\footnotetext{dong.wang@yizhun-ai.com}
\stepcounter{footnote}\footnotetext{tbh691203@163.com}%
\stepcounter{footnote}\footnotetext{hlbcus@163.com}%
\stepcounter{footnote}\footnotetext{zhuqingli@pumch.cn}
\stepcounter{footnote}\footnotetext{wangyong@cicams.ac.cn}%
\stepcounter{footnote}\footnotetext{wanglw@pku.edu.cn}

\begin{figure*}[t]
    \centering
    \includegraphics[width=\textwidth]{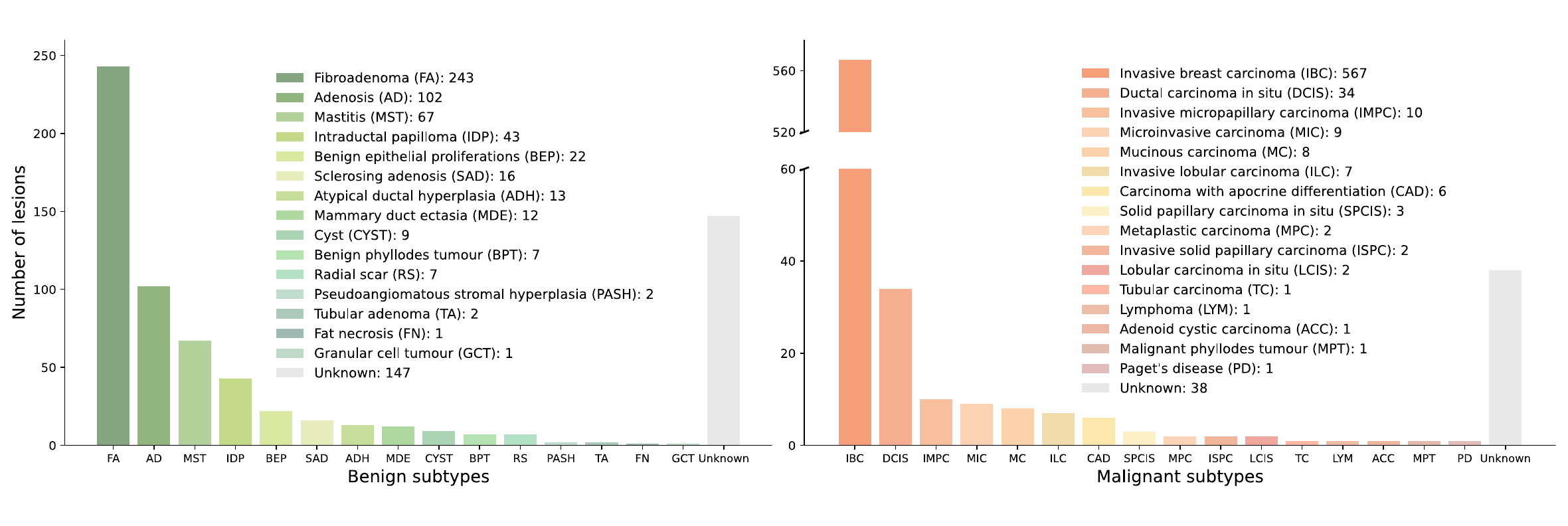}
    \caption{\small \small {\bf The challenge of long-tail distribution.} The distribution of pathological subtypes is long-tailed in our training set which has 1,387 biopsy-confirmed lesions. In benign lesions, the two most frequent subtypes together account for 49.7\% of the lesions, with the remaining 13 subtypes comprising 50.3\%. In malignant lesions, the most frequent subtype accounts for 81.8\% of the lesions, while the remaining 15 subtypes comprise only 18.2\%.
    }
    \label{fig:problem}
\end{figure*}

Breast cancer has become the most common cancer among women globally~\cite{siegel2023cancer,chhikara2023global,xia2022cancer}, and early detection can significantly decrease the mortality rates~\cite{zielonke2020evidence}.
In breast cancer detection, ultrasound (US) is an essential imaging method widely adopted worldwide for its safety and low cost~\cite{sickles2013acr,shen2015multi,sood2019ultrasound}.
Accurately interpreting breast-US findings poses a great challenge~\cite{lazarus2006bi} as it requires radiological knowledge to comprehensively analyze clinically relevant features~\cite{sickles2013acr} such as margin characteristics, echo patterns, shape, and calcifications. Data-driven deep learning models provide a promising solution for accurate breast-US diagnoses~\cite{shen2021artificial,qian2021prospective,ng2023prospective}. However, the collected training data~\cite{shen2021artificial,qian2021prospective,lin2022new,al2020dataset,ng2023prospective} is often limited and inherently exhibits a long-tail distribution of pathological subtypes~\cite{ohuchi2016sensitivity,berg2016ultrasound,zhao2022enhancing,tan20202019}, as shown in \Cref{fig:problem}. When learning from such limited and imbalanced data, the models tend to primarily focus on predicting the head categories correctly, making it more likely to produce wrong predictions for rare categories~\cite{zhang2023deep}. Moreover, rare categories can be error-prone for radiologists, particularly requiring AI assistance. Notably, specifically collecting sufficient tail data can be extremely costly due to the rarity of these cases, not to mention the issues associated with medical data collection, such as privacy concerns, high costs, and legal risks.

\begin{figure*}[t]
    \centering
    \includegraphics[width=\textwidth]{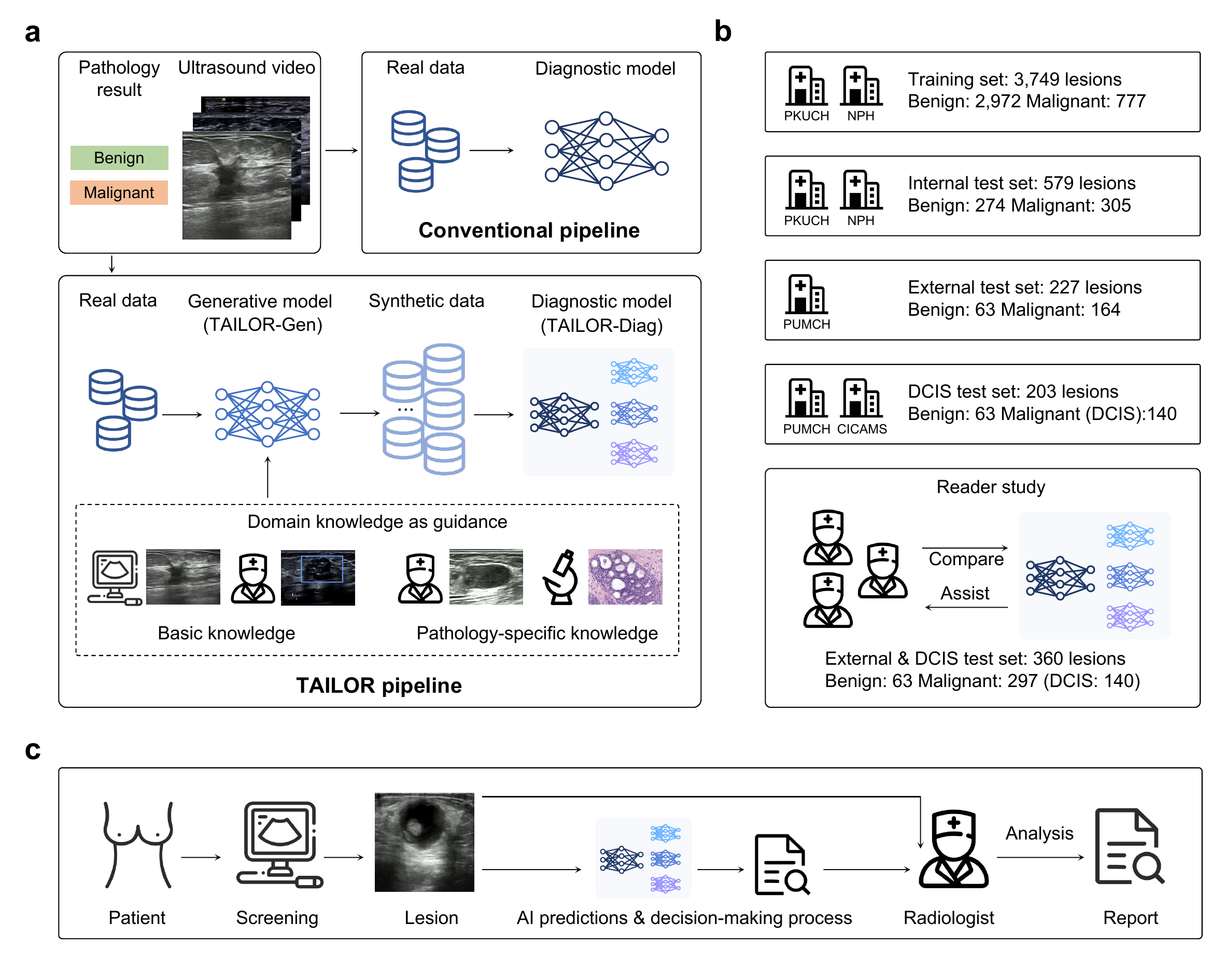}
    \caption{\small \small {\bf Overview of TAILOR and our study design.} {\bf a}, TAILOR pipeline vs. conventional pipeline. TAILOR utilizes knowledge-driven AI-generated data for accurate and interpretable diagnoses. {\bf b}, Study design. The number of lesions in the training set, the internal test set, the external test set, and the DCIS test set. The design of reader study. The involved four institutions are introduced in \Cref{subsec:data_process}. {\bf c}, AI-assisted clinical diagnosis. We compared TAILOR-Diag with radiologists. We investigate the effectiveness of the TAILOR-Diag's assistance to enhance radiologists' diagnostic performance.
    }
    \label{fig:workflow}
\end{figure*}

Recent advances in generative models~\cite{ho2020denoising,ouyang2022training,rombach2022high,neil2023synthetic,gao2023synthetic} have made it possible to produce realistic and diverse content according to the input instructions or conditions. 
Moreover, these models demonstrate notable transferability: with only a small amount of domain-specific data, they can be efficiently fine-tuned to generate high-quality outputs tailored to targeted scenarios~\cite{ouyang2022training,hu2022lora,ruiz2023dreambooth}. Given these successes, we propose TAILOR, a pipeline that trains an accurate and interpretable diagnostic model (TAILOR-Diag) with the help of a knowledge-driven generative model (TAILOR-Gen), as illustrated in \Cref{fig:workflow}.

To briefly introduce, we first train a diffusion generative model, TAILOR-Gen, to generate knowledge-conditioned images. Besides knowledge of benign and malignant pathology, we incorporate critical domain knowledge including various information, such as rare pathological subtypes, error-prone US features, and visual imaging appearances, which we observe have limited diversity in the training data. The annotations for the knowledge information come from pathology results, US reports, or expert opinions. By incorporating proper knowledge, the model can learn and generate images conditioned on more contexts, significantly improving the quality of the generated images, especially for rare categories. With the trained TAILOR-Gen, we generate large-scale, diverse, and realistic data, and build the diagnostic model TAILOR-Diag using the synthetic dataset (\Cref{fig:workflow}a). In particular, we design TAILOR-Diag as an ensemble of multiple classifiers that adaptively leverage appropriate knowledge to accurately classify the benign and malignant pathology. Therefore, the decision-making process of the model is interpretable and understandable for human users~\cite{guyatt1993users}.

Extensive results demonstrate that TAILOR facilitates accurate and interpretable breast-US diagnoses. In terms of accuracy, TAILOR-Diag (AUC=0.954, 95\% Confidence Interval (CI) 0.932$-$0.983) outperforms the baseline trained on real data (AUC=0.909, 95\% CI 0.867$-$0.947) on the external test set, significantly improving the performance to exceed the average performance of nine board-certified breast-US radiologists by 33.5\% (95\% CI 23.2$-$44.1\%) in specificity with the same sensitivity. Moreover, in diagnoses of ductal carcinoma in situ (DCIS), an error-prone subtype of early-stage cancer, TAILOR-Diag outperforms all nine radiologists by a large margin, with only 34 DCIS cases in the source data. In terms of interpretability, we investigate whether the assistance of TAILOR-Diag can improve radiologists' performance in real clinical settings. Notably, the average performance of nine radiologists improves by 6.4\% (95\% CI 3.8$-$8.9\%) in specificity without loss of average sensitivity. These impressive results demonstrate that our proposed pipeline, TAILOR, can effectively learn critical knowledge from a small amount of domain-specific data, which has the potential to be extended to various diseases and imaging modalities.

\section{Results}
\label{sec:result}

\subsection{Datasets}
\label{subsec:overview}

In this work, we conducted a multi-centre study with US images of breast lesions recruited from four institutions in China (\Cref{fig:workflow}b).
The involved institutions enable us to collect data from representative patient populations, detailed in \spsection{1.1}. For training and internal evaluation, we retrospectively collected scanning videos of 3,422 patients with 4,328 lesions from two internal institutions and split the internal dataset by patients. The training set consisted of 3,749 lesions (1,387 biopsy-confirmed lesions), and the internal test set comprised 579 biopsy-confirmed lesions. After we developed TAILOR-Diag, for external evaluation, we prospectively collected breast-US images of 225 consecutive patients with 227 biopsy-confirmed lesions from an external institution. To accurately evaluate the model performance on DCIS, we purposely collected 133 biopsy-confirmed DCIS lesions from two external institutions. We conducted the reader study on the (random shuffled) mixed test set of the external consecutive lesions and the purposely collected DCIS lesions where nine radiologists interpreted these lesions and attempted to integrate TAILOR-Diag into the clinical workflow (\Cref{fig:workflow}c). The details of dataset construction are described in Section~\ref{subsec:data_process} and the patient demographics and lesion characteristics are illustrated in \sptable{2}.

\subsection{Knowledge-driven generative model}
\label{subsec:knowledge}

\begin{figure*}[t]
    \centering
    \includegraphics[width=0.98\textwidth]{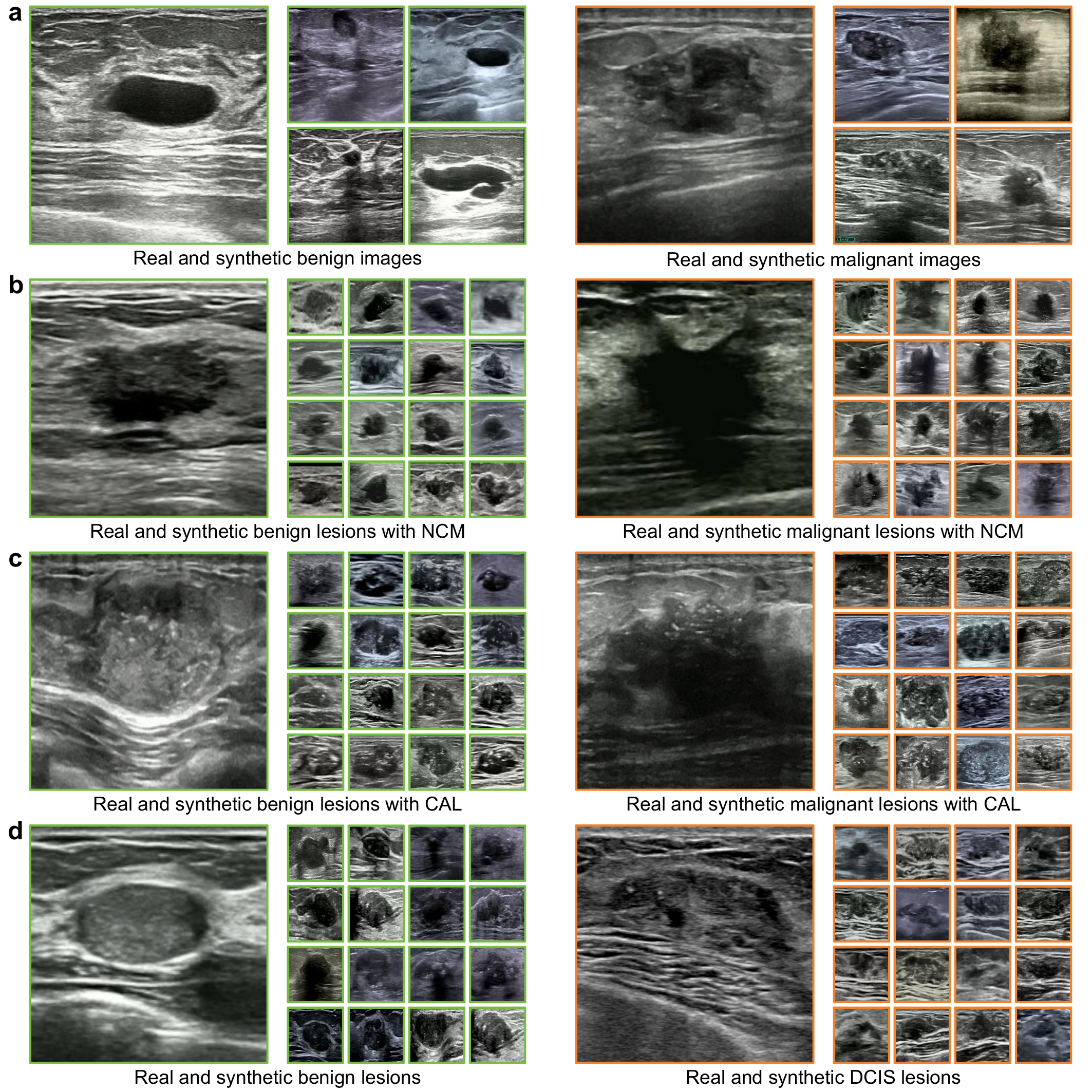}
    \caption{\small \small {\bf Visualization of real and synthetic breast-US data.}
    Real and synthetic lesions for the pathology classification tasks. {\bf a}, Images with common benign and malignant lesions. {\bf b}, Benign and malignant lesions with NCM. {\bf c}, Benign and malignant lesions with CAL. {\bf d}, Benign and DCIS lesions. The large images are collected real data, and the smaller images are synthetic data produced by TAILOR-Gen. To demonstrate the realism of the lesion and background areas in the generated images, we provide the whole-slide synthetic images in {\bf a}. To demonstrate the representative US features of each tail category, we provide the lesion areas of the generated images in {\bf b}, {\bf c}, and {\bf d}.
    }
    \label{fig:vis}
\end{figure*}

\begin{figure*}[t]
    \centering
    \includegraphics[width=\textwidth]{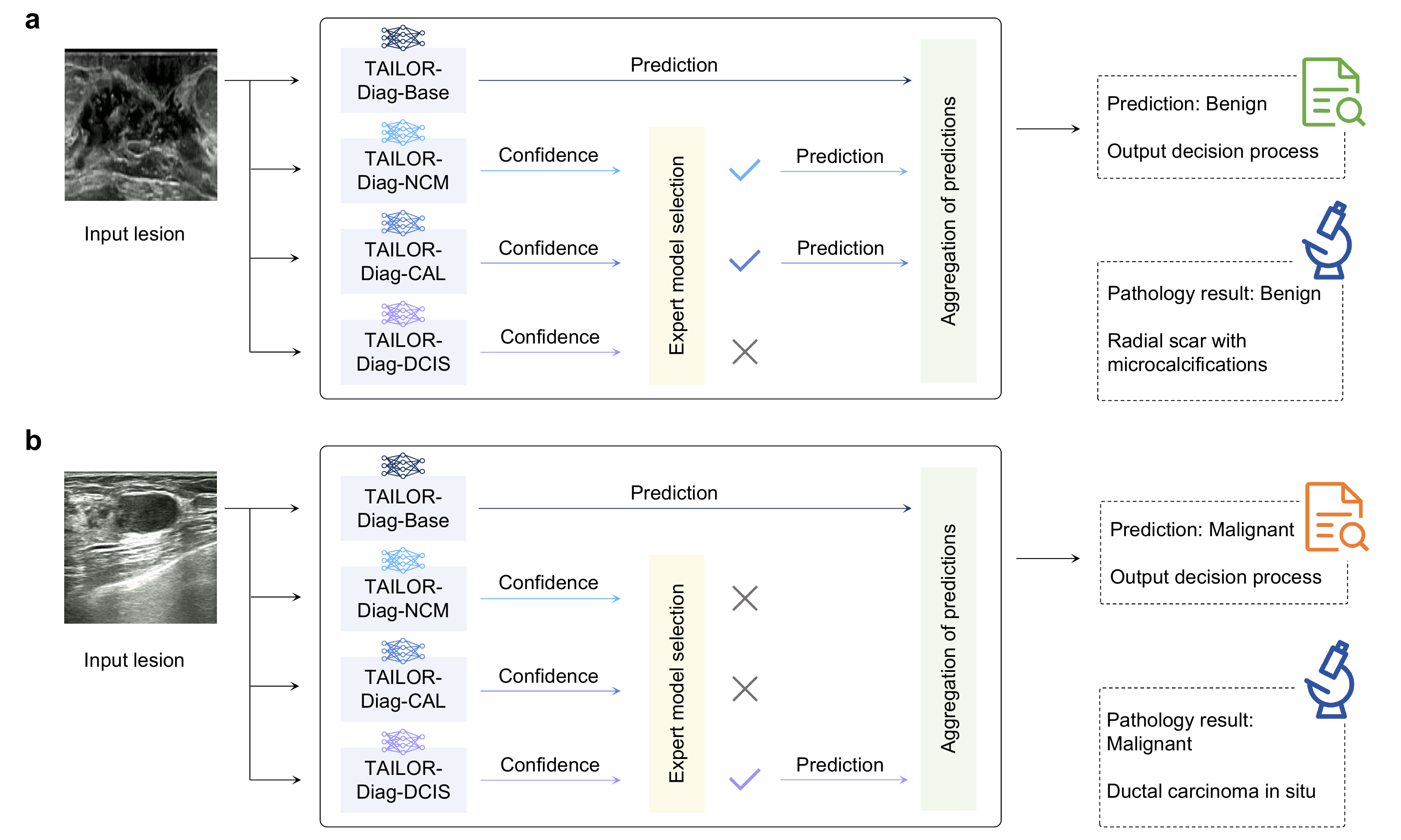}
    \caption{\small \small {\bf Interpretable diagnostic model.}
    We provide two examples of TAILOR-Diag's decision-making processes (for {\bf a}, a benign lesion with NCM and CAL, {\bf b}, a malignant DCIS lesion). For each input image, it passes through the general model and automatically selects expert models based on confidence scores. Then, we combine the predictions of both the general model and the selected expert model(s) to obtain the final prediction.
    }
    \label{fig:tailor_inference}
\end{figure*}

We seek to develop an accurate and interpretable deep learning model for breast-US diagnoses. To achieve this goal, we propose to augment the limited and long-tailed data using a knowledge-driven generative model called TAILOR-Gen. Specifically, TAILOR-Gen targets learning both the basic knowledge and the pathology-specific knowledge under expert supervision. Basic knowledge is useful for enhancing the diversity of visual appearance. More importantly, the pathology-specific knowledge is critical for accurate diagnoses.

We define the basic knowledge as the visual appearances of factors not strongly correlated with lesion pathology, varying across different patients and clinical situations. Diagnostic models might learn incorrect correlations between these factors and lesion pathology when trained on a dataset with limited diversity in visual appearances, which poses challenges in model generalization to different clinical situations. Here, we explore the basic knowledge of lesion area and device type to enrich the data diversity. The lesion area refers to the relative position and scale of lesions on US screens which could vary as radiologists adjust them for different diagnostic purposes. Additionally, device types can introduce variations in image quality, texture, or color bias. More details are provided in  \spfigure{3} and \spfigure{4}. 

The pathology-specific knowledge establishes the connections between US features and lesion pathology, thus being critical for accurate diagnoses. However, generative models trained directly on binary pathology labels tend to learn knowledge from head categories. In this study, TAILOR-Gen is designed to learn the pathology-specific knowledge for both head and tail categories. To identify underrepresented tail categories, we investigate the US features and the pathological subtypes. First, we investigate US features, defined in the American College of Radiology published Breast Imaging Reporting and Data System (BI-RADS) lexicon guidelines~\cite{sickles2013acr}. In clinical practice, US features are evaluated by radiologists based on US images and their experience. Different US features can indicate different probabilities of malignancy. Here, we first explore two critical US features.
\begin{itemize}
\small
    \item Not circumscribed margins (NCM) refer to the unclear boundary between lesions and surrounding tissues. NCM often suggests malignant breast cancer, while some rare benign lesions can also exhibit NCM~\cite{kim2008correlation,ko2010triple,wojcinski2012sonographic,su2017non}, such as radial scar and mastitis.
    \item Microcalcifications in a mass (CAL) are calcium deposits $<0.5$ mm in diameter embedded in a mass, recognized as small hyperechoic foci in US images. CAL often appears in breast cancer, while sometimes they can also be found in benign lesions~\cite{sickles1986breast,tse2008calcification,demetri2012breast,logullo2022breast}.
\end{itemize}
Thus, benign lesions with US features of NCM or CAL are two tail categories that can be challenging in clinical practice. Second, for pathological subtypes, we reference the taxonomy defined in the WHO classification~\cite{tan20202019} and other professional books on pathology~\cite{hoda2014rosen,peng2019practical}. In clinical practice, pathological subtypes are determined by surgery or biopsy, reflecting cellular-level lesion structures. Note that the pathological subtypes can not be determined by radiologists directly from US images. Here, we investigate an error-prone pathological subtype that is critical in the early detection of breast cancer.
\begin{itemize}
\small
    \item Ductal carcinoma in situ (DCIS) is a non-invasive early-stage pathological subtype where all cancer cells are confined within the basement membrane~\cite{pinder2010ductal,ernster1997increases,winchester2000diagnosis}. DCIS lacks typical malignant features of invasive cancer and sometimes exhibits non-mass lesions or nodules with regular shape or circumscribed margins. These features may be associated with benign findings in clinical practice.
\end{itemize}

We train a generative model, TAILOR-Gen, to learn the aforementioned knowledge, enabling it to produce realistic and diverse data that encompasses this knowledge. Specifically, TAILOR-Gen is designed as a conditional Denoising Diffusion Probabilistic Model (DDPM)~\cite{ho2020denoising,dhariwal2021diffusion,ho2021classifierfree} that can produce images according to input conditions.
First, we pre-train TAILOR-Gen on the entire training set conditioned on the benign or malignant pathology labels, enabling it to generate images based on pathology conditions. Therefore, these pathology conditions can be used as pseudo-labels to train diagnostic models. With the powerful DDPM, the generated images contain realistic lesions and background areas: lesions can accurately reflect the representative US features of the given pathology, and the background areas accurately reflect the structures and textures of breast anatomy such as skin, fat, and gland tissue. Second, we fine-tune TAILOR-Gen to incorporate the basic knowledge. We annotate the lesion bounding boxes and device types by radiologists based on the US reports. Then, we fine-tune TAILOR-Gen conditioned on these new annotations, enabling it to produce customized breast-US images with specific lesion areas and device types controlled by the conditioning inputs. Specifically, we sample up to 800,000 breast-US images that are well-balanced in pathology and diverse in visual appearance (\Cref{fig:vis}a). Third, we fine-tune TAILOR-Gen conditioned on pathology-specific labels. To incorporate pathology-specific knowledge, we annotate NCM and CAL labels with expert guidance and identify DCIS lesions based on pathology results. Leveraging the notable transferability of DDPM, we fine-tune TAILOR-Gen conditioned on these specifically annotated lesions, enabling it to produce images for each tail category. Then, we sample 100,000 images encompassing critical knowledge for each of the three tail categories. Specifically, we sample NCM or CAL lesions with balanced pathology labels (\Cref{fig:vis}b, c), as well as balanced DCIS and benign lesions (\Cref{fig:vis}d). It is important to note that different conditions can be combined to guide TAILOR-Gen to produce images for tail categories with diverse visual appearances.

\subsection{Interpretable diagnostic model}
\label{subsec:expert}

Using TAILOR-Gen, we manage to generate diverse and well-balanced data, which can be used to improve the training of the diagnostic model. Based on the generated data, we train a diagnostic model, TAILOR-Diag, to learn critical domain knowledge for accurate diagnoses. We design TAILOR-Diag as an ensemble of four classification models to diagnose lesions with proper knowledge: a general model primarily for head categories and three expert models for each of the three tail categories. Each classifier consists of a Swin Transformer~\cite{liu2021swin} backbone, and a binary classification head to predict pathology categories.

To optimize the general predictive ability of common cases, we pre-train a classification model on the aforementioned 800,000 generated images, called TAILOR-Diag-Base. The generated images offer a broader visual variety than conventional data augmentations, enabling the classifier to better generalize to different clinical situations. After the pre-training finishes, we fine-tune TAILOR-Diag-Base using 100,000 tailored images for each tail category to enhance the specialized predictive ability. These expert models focus on different aspects and provide confidence scores of their predictions, named TAILOR-Diag-NCM, -CAL, and -DCIS respectively. As expert models specifically learn critical knowledge of rare cases, we believe they improve the general model's predictive ability of tail categories. Therefore, combining the predictions of both general and expert models is expected to yield better results in real clinical settings. To leverage the expertise of each model, we design a decision-making process wherein an input image passes through the general model and automatically selects expert models based on their confidence scores and then combines both the general and the expert model predictions to obtain the final prediction. Note that images with common lesions may not select any expert models. We provide two examples of the decision-making process of TAILOR-Diag in \Cref{fig:tailor_inference}.

TAILOR-Diag is interpretable and understandable by radiologists because its decision-making process mimics the diagnostic strategies used by human experts in the real-world clinical diagnosis process~\cite{guyatt1993users}. For typical common cases, the general model provides predictions that are often consistent with radiologists' opinions, thereby enhancing their confidence. For uncertain rare cases, radiologists can carefully analyze the predictions of expert models. This detailed analysis allows radiologists to revise their initial diagnoses based on the prediction results from the model, or correct the model's errors based on their knowledge, ultimately leading to more accurate diagnoses. In \Cref{fig:tailor_inference}a, for the challenging benign lesion with both NCM and CAL, the expert models TAILOR-Diag-NCM and -CAL are selected and predict a high probability of benignity; and in \Cref{fig:tailor_inference}b, for the challenging DCIS lesion, the expert model TAILOR-Diag-DCIS is selected and predict high probability of malignancy. These hints enable radiologists to revise their predictions for more accurate diagnoses.

\subsection{General evaluation}
\label{subsec:evaluation}

\begin{figure*}[t]
    \centering
    \includegraphics[width=\textwidth]{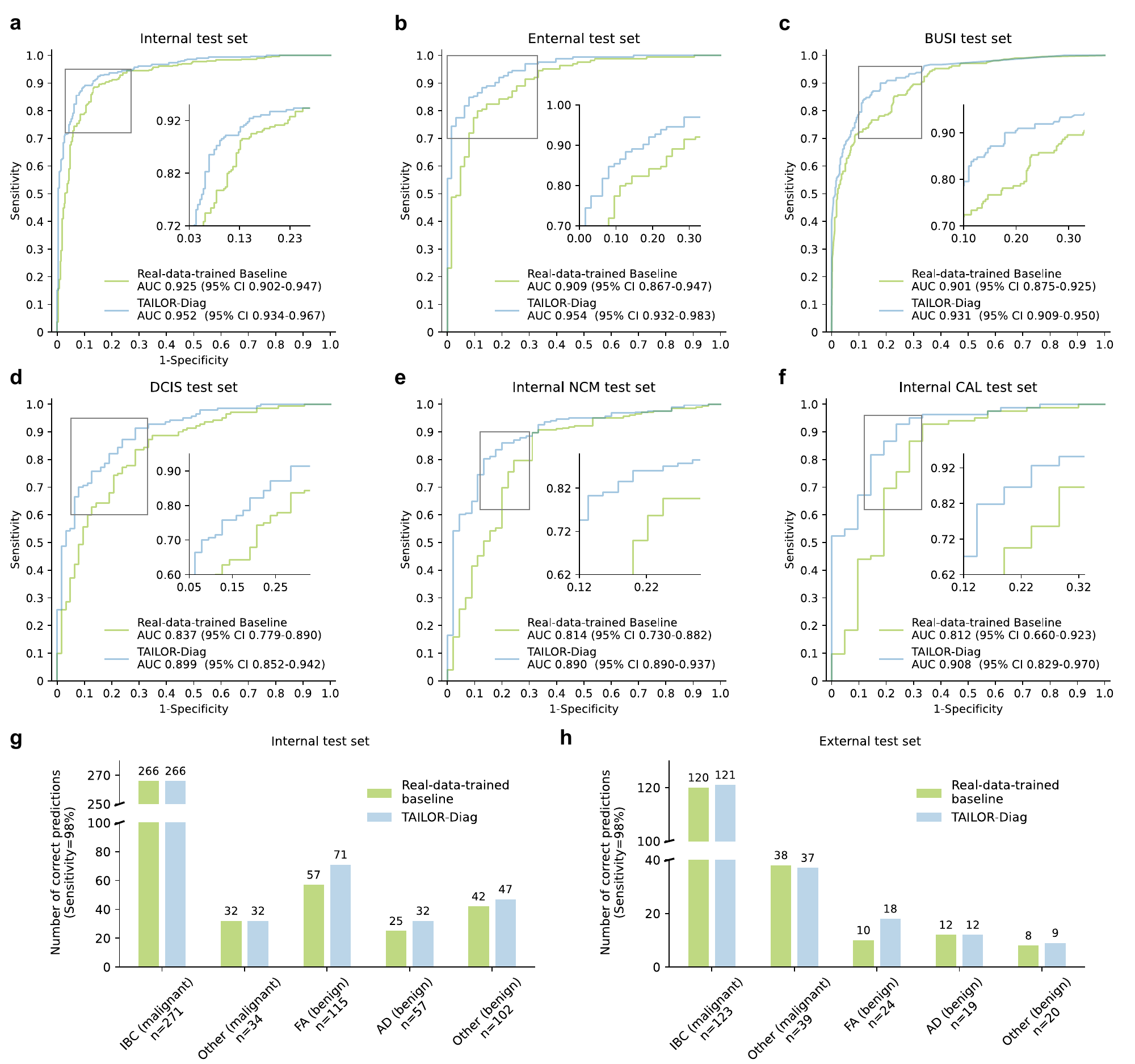}
    \caption{\small \small {\bf Comparison of the real-data-trained baseline and TAILOR-Diag.}  
    We show the receiver operating characteristic (ROC) curves on {\bf a}, the internal test set, {\bf b}, the external test set, and {\bf c}, the public BUSI test set.
    We show the ROCs of the pathology classification task on {\bf d}, DCIS and benign lesions, {\bf e}, lesions with NCM and {\bf f}, lesions with CAL.
    We provide the number of correct predictions for each pathological subtype on {\bf g}, the internal test set, and {\bf h}, the external test set. The results of the real-data-trained baseline and TAILOR-Diag are both calculated with a fixed sensitivity of 98\%.
    }
    \label{fig:real_synt}
\end{figure*}

\begin{figure*}[t]
    \centering
    \includegraphics[width=\textwidth]{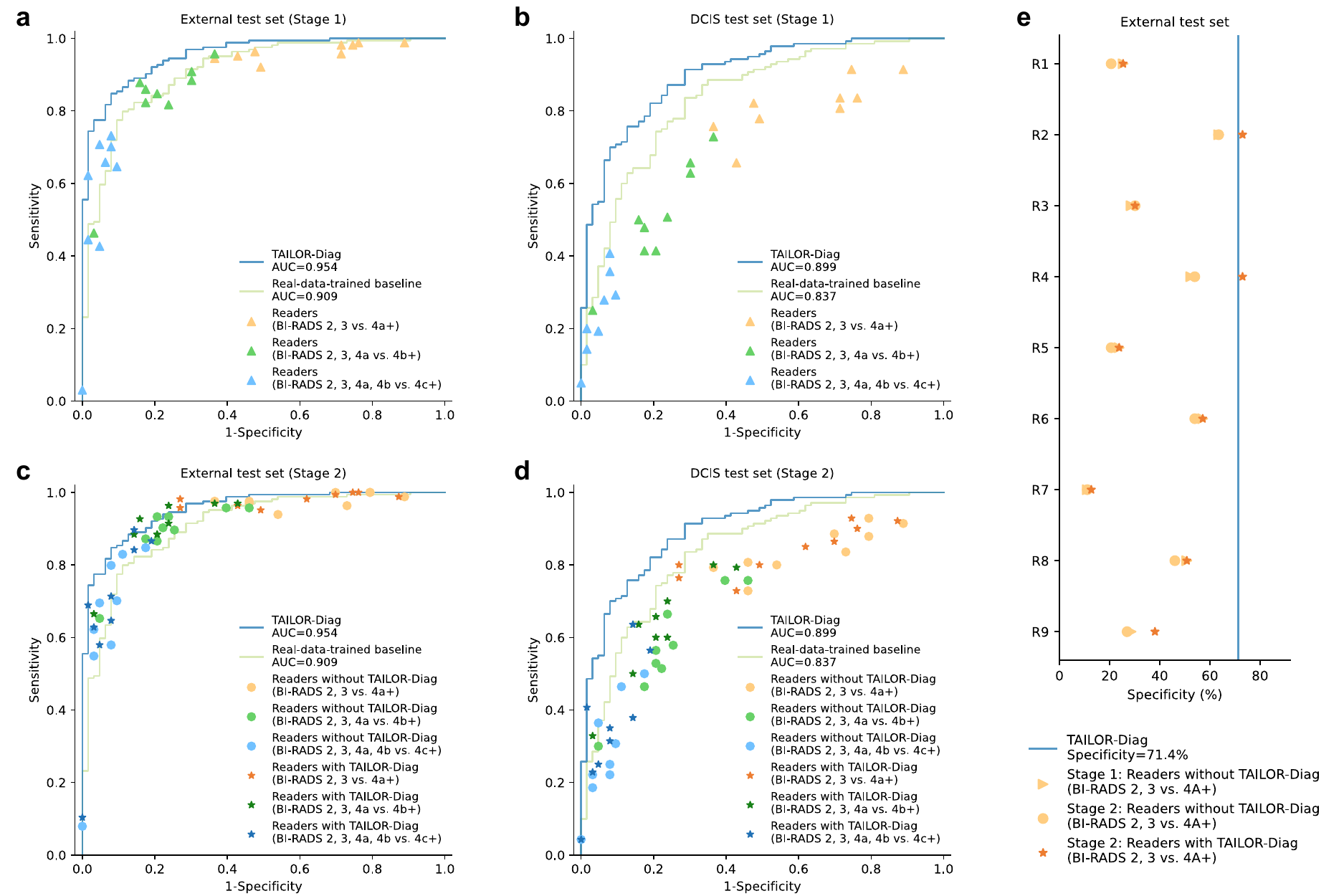}
    \caption{\small \small {\bf Reader study results.} ROC curves of TAILOR-Diag and readers results (with thresholds BI-RADS 4A, 4B, and 4C) are shown in {\bf a$-$d}.
    As shown in {\bf a} and {\bf c}, in the first stage, TAILOR-Diag outperforms readers at different thresholds on external and DCIS test sets using only B-mode images.
    As shown in {\bf b} and {\bf d}, in the second stage, readers improve with the assistance of TAILOR-Diag on the external and DCIS test sets under real clinical settings.
    {\bf e}, With the assistance of TAILOR-Diag, readers achieve consistent improvements in specificity on the external test set without loss of sensitivity. The results are calculated with the threshold of BI-RADS 4A.
    }
    \label{fig:reader_study}
\end{figure*}

We first evaluate TAILOR-Diag on the internal test set, which consists of 579 lesions (274 benign and 305 malignant). All lesions have biopsy-confirmed pathology results, called ``gold standard" labels. To demonstrate the strength of using generated data, we compare TAILOR with the conventional pipeline. In the conventional pipeline, we train a diagnostic model with the same classifier architecture (without the decision-making process) on the collected training set with resampling techniques to re-balance the pathology categories, named real-data-trained baseline. On the internal test set, TAILOR-Diag achieves an area under the receiver operating characteristic curve (AUC) of 0.952 (95\% CI 0.934$-$0.967). For comparison, the real-data-trained baseline only achieves an AUC of 0.925 (95\% CI 0.902$-$0.947, P-value=0.0001). We plot the receiver operating characteristic (ROC) curves of both models in \Cref{fig:real_synt}a. These results demonstrate that our TAILOR pipeline facilitates significantly better diagnostic performance than the conventional pipeline.

To further evaluate the model's ability, we test it on the datasets from external institutions with various patient populations and imaging protocols. First, we assess the models on the prospective consecutive external test set consisting of 227 lesions (63 benign and 164 malignant) with ``gold standard" labels. On this task, TAILOR-Diag achieves an AUC of 0.954 (95\% CI 0.932$-$0.983) while the real-data-trained baseline only achieved an AUC of 0.909 (95\% CI 0.867$-$0.947, P-value=0.0023), as shown in \Cref{fig:real_synt}b.
Second, we evaluate the trained models on a public Breast Ultrasound Images (BUSI) dataset~\cite{al2020dataset} collected from an institution in Egypt (437 benign, 210 malignant, and 133 negative lesions) where negative lesions are not used in our evaluation. TAILOR-Diag achieves an AUC of 0.931 (95\% CI 0.909$-$0.950) while the real-data-trained baseline achieved an AUC 0f 0.901 (95\% CI 0.875$-$0.925, P-value=0.0001), as shown in \Cref{fig:real_synt}c.
All these results demonstrate that TAILOR-Diag has great generalization ability, achieving significantly better performance than the real-data-trained baseline.

\subsection{Fine-grained evaluation on specific categories}
\label{subsec:tail}

In this subsection, we give a detailed analysis of the model performance on specific categories. First, we focus on the three investigated error-prone tail categories. For DCIS, we calculate the pathology prediction of different models on the DCIS test set consisting of 63 benign lesions (from the external test set) and 140 DCIS lesions (7 DCIS lesions are from the external test set and 133 DCIS lesions are additionally collected). As shown in \Cref{fig:real_synt}d, TAILOR-Diag achieves an AUC of 0.899 (95\% CI 0.852$-$0.942) while the real-data-trained baseline achieves an AUC of 0.837 (95\% CI 0.779$-$0.890, P-value=0.0026). For NCM, with expert guidance, we annotate 324 lesions with NCM (45 benign and 279 malignant) in the internal test set. As shown in \Cref{fig:real_synt}e, on lesions with NCM, TAILOR-Diag achieves an AUC of 0.890 (95\% CI 0.835$-$0.937) while the real-data-trained baseline achieves an AUC of 0.814 (95\% CI 0.730$-$0.882, P-value=0.0008). For CAL, we annotate 103 lesions with CAL (21 benign and 82 malignant) in the internal test set. As shown in \Cref{fig:real_synt}f, on lesions with CAL, TAILOR-Diag achieves an AUC of 0.908 (95\% CI 0.829$-$0.970) while the real-data-trained baseline achieves an AUC of 0.812 (95\% CI 0.660$-$0.923, P-value=0.0085). All of the results show the superiority of TAILOR-Diag compared to the conventional approach.

Second, we assess whether the performance gains are consistent across pathological subtypes. We report the number of correct predictions for each pathological subtype when the overall sensitivity is 98\% (radiologists achieved an average sensitivity of 97.9\% under the real clinical settings in \Cref{subsec:reader_study}). Specifically, we evaluate the performance of different pathological subtypes, including invasive breast carcinoma (IBC), fibroadenoma (FA), and adenosis (AD). Other subtypes are combined due to the small number of lesions. 
TAILOR-Diag demonstrates comparable results to the real-data-trained baseline on malignant subtypes, maintaining the same sensitivity.
For benign subtypes, TAILOR-Diag consistently outperforms the baseline across subtypes on both internal and external test sets, as shown in \Cref{fig:real_synt}g and h.

\subsection{Reader study}
\label{subsec:reader_study}

To further demonstrate the strength of TAILOR-Diag, we conducted a reader study to compare the models with human radiologists and investigate how TAILOR-Diag can assist radiologists in practice. Here, we used the mixed test set with 227 consecutive lesions and the purposely collected 133 DCIS lesions. We invited nine board-certified breast-US radiologists with a range of experience of 3$-$26 years (11 years on average) to analyze these lesions and provide their predicted BI-RADS scores. Because the distribution of data used in this study differs from that in clinical practice, we specifically informed readers that they should independently evaluate each lesion. We calculated the sensitivity and specificity of readers using the BI-RADS 4A as the threshold for determining the binary predictions (BI-RADS 2, 3 as benignity, and BI-RADS 4A$+$ as malignancy).

The reader study consisted of two stages. 
In the first stage (Stage 1), we provided the B-mode breast-US images to both TAILOR-Diag and the readers and compared their predictions. As shown in \Cref{fig:reader_study}a, TAILOR-Diag consistently outperformed nine readers on different BI-RADS thresholds. 
On the 227 consecutive lesions, TAILOR-Diag outperformed the average reader performance by 33.5\% (95\% CI 23.2$-$44.1\%, P-value=0.0002) in specificity with the same sensitivity of 96.4\%, and outperformed the average reader performance by 3.0\% (95\% CI 1.4$-$4.8\%, P-value=0.0022) in sensitivity (TAILOR-Diag achieved 99.4\% (95\% CI 93.1$-$100\%)) with the same specificity of 37.9\%. The ROCs of TAILOR-Diag, the real-data-trained baseline, and the results of readers on the external test set are shown in \Cref{fig:reader_study}a. For diagnoses of DCIS, TAILOR-Diag outperformed the mean performance of readers by 43.0\% (95\% CI 31.9$-$53.6\%, P-value$<$0.0001) in specificity with the same sensitivity of 81.3\%, and outperformed the average reader performance by 16.5\% (95\% CI 11.5$-$21.4\%, P-value$<$0.0001) in sensitivity with the same specificity of 37.9\%. The ROCs of TAILOR-Diag, the real-data-trained baseline, and the results of readers on the DCIS test set are shown in \Cref{fig:reader_study}b. These results demonstrate that TAILOR-Diag is more accurate than human radiologists using the same input information of B-mode images.

The second stage (Stage 2) evaluated the effectiveness of the TAILOR-Diag's assistance for radiologists in real clinical settings. To mimic practical conditions, besides B-mode images, we further provided readers with patient demographics and color Doppler images, then required readers to re-assess the lesions with this additional information. The results demonstrate that with the help of this information, the average reader performance did not significantly change. The average reader sensitivity (97.9\%) improved by 1.5\% (95\% CI 0.4$-$2.8\%), but the average reader specificity decreased by 1.6\% (95\% CI -1.6$-$4.7\%) compared to the results in Stage 1. Next, we provided readers with TAILOT-Diag predictions and decision-making processes and required them to re-assess the lesions with AI assistance. The performance gains of each reader in Stage 2 using different BI-RADS scores as thresholds are shown in \Cref{fig:reader_study}c and \Cref{fig:reader_study}d. With the assistance of TAILOR-Diag, the average reader performance improved by 6.4\% (95\% CI 3.8$-$8.9\%) in specificity without loss of sensitivity (improved by 0.1\%) on the external test set, as shown in \Cref{fig:reader_study}e. Moreover, two human radiologists exceeded the performance of TAILOR-Diag with its assistance. They not only revised their misdiagnoses with the model's hints but also pointed out the model's error based on their analysis of the decision-making processes, proving the notable interpretability of TAILOR-Diag. These results demonstrate that incorporating TAILOR-Diag into the clinical workflow can improve the diagnostic performance of radiologists, especially in specificity, under real clinical settings. More details of the reader study are illustrated in \spsection{4}. 

\section{Discussion}
\label{sec:discussion}

Data-driven deep learning models have demonstrated significant capabilities in assisting radiologists with diagnosing a wide range of diseases across various imaging modalities~\cite{de2018clinically,ardila2019end,ouyang2020video,mckinney2020international,shen2021artificial,qian2021prospective,cao2023large,zhou2023foundation}. The success of these diagnostic models is largely attributed to high-quality datasets that encompass rich domain knowledge essential for clinical diagnoses. However, medical data collection faces challenges due to privacy, cost, and legal issues, leading to limitations in source datasets~\cite{price2019privacy,malin2013biomedical,jones2015systematic}. To address these challenges, most previous works explored the use of generative models as a way for data augmentation~\cite{antoniou2017data,mariani2018bagan,frid2018gan,sandfort2019data,gupta2019generative,ghorbani2020dermgan,xue2021selective,chambon2022adapting,sun2023aligning,hu2023label,sagers2023augmenting,pinaya2023generative,ktena2024generative}. In this study, we take a step back to rethink the synthetic data augmentation methods and find that incorporating domain knowledge into the synthetic data is more important (\spsection{3.1}).
 
We leverage the recent advances in generative models~\cite{ouyang2022training,ruiz2023dreambooth} to produce high-quality data for rare lesions using the long-tailed medical dataset.
In computer vision, techniques of generative models have been developed to address a similar challenge. Previous works demonstrate that a pre-trained conditional generative model can be ``personalized" to produce photos of a specific person by learning shared knowledge from the entire dataset and learning identification knowledge from 3$-$5 photos~\cite{ruiz2023dreambooth,wang2024instantid}. 
With this insight, we follow the same way and develop a knowledge-driven generative model in the medical domain that learns the basic knowledge from the whole dataset and the pathology-specific knowledge from a few tail-category lesions. Leveraging these capabilities, our synthetic breast-US images demonstrate realism and diversity, proving useful in downstream tasks.

Our study has the potential for application in practical clinical scenarios. For breast cancer early detection, TAILOR-Diag significantly outperforms human radiologists on DCIS, a critical subtype of early-stage cancer. This makes it suitable for integration into the breast screening workflow. Additionally, TAILOR-Diag can be used to re-evaluate retrospective breast-US examinations. As a high-throughput method, TAILOR-Diag can re-evaluate large-scale preserved breast-US data in hospitals, identifying potential false negatives and prompting further examinations. These improvements can contribute to better treatment outcomes and reduced mortality rates.

The proposed TAILOR pipeline offers promising future directions for exploration. First, integrating multi-modal breast-US inputs, such as color Doppler, elastography US, and dynamic video information, could further improve diagnostic performance~\cite{qian2021prospective}. Second, besides the three tail categories investigated in this study, TAILOR can be adapted to incorporate domain knowledge for other error-prone categories, potentially further enhancing breast-US diagnostic performance. Finally, we believe that TAILOR can be extended to various diseases and imaging modalities beyond breast-US diagnoses.

\section{Methods}
\label{sec:methods}

\subsection{Ethical approval}

Our study was approved by the institutional review board of the Peking University Cancer Hospital \& Institute (ID: 2024YJZ41). The study was not interventional and was performed under guidelines approved by the institutional review board. Informed consent was waived since the study presents no more than minimal risk. All datasets processed for this research were de-identified before transfer to study investigators.

\subsection{Breast-US data acquisition, processing, and annotation}
\label{subsec:data_process}

To conduct the multi-centre study, we collected data from four Grade-3A hospitals in China: Peking University Cancer Hospital \& Institute (PKUCH), Nanchang People's Hospital (NPH), Peking Union Medical College Hospital (PUMCH) and Cancer Institute, Chinese Academy of Medical Sciences (CICAMS). We defined two hospitals, PKUCH and NPH, as internal institutions where we collected data for training and internal evaluation; and the other two hospitals, PUMCH and CICAMS, were defined as external institutions where we collected data for external evaluation.

We collected breast-US scanning videos as the internal dataset and then divided them into a training set and an internal test set. Here, we regarded videos as sequential 2D images, as we used the image generative models. The videos were collected from patients who underwent breast-US examinations at PKUCH and NPH between January 2020 and March 2021. We collected US videos instead of US images preserved in the standard clinical workflow because videos contained continuous frames in scanning processes, offering more information than discrete images to train generative models. In data processing, we retained B-mode US frames that clearly showed lesions without blurring in the lesion-scanning process, excluding frames in the initial lesion-finding process. Following the standard workflow~\cite{qian2021prospective}, when multiple lesions were detected in a breast, only the major lesion was included. As detailed in \spsection{1.3}, radiologists annotated lesion areas using bounding boxes, and device types were extracted from the US reports.
In the training set, we kept video clips of 3,749 lesions (2,972 benign and 777 malignant) after pre-processing, consisting of 2,589,824 frames (1,905,670 benign and 684,154 malignant). Note that these frames contain redundant temporal information with limited diversity in visual appearance. Out of these 3,749 lesions, 1,387 lesions (694 benign and 693 malignant) had biopsy-confirmed pathology results, serving as ``gold standard" labels. The remaining 2,362 lesions were assigned ``silver standard" pathology labels under the expert guidance, based on BI-RADS scores~\cite{sickles2013acr}. Specifically, lesions with BI-RADS 2 or 3 were labeled as benign, those with BI-RADS 4C or higher as malignant, and the others were excluded. The retained 2,362 lesions all received ``silver standard" labels of benign or malignant pathology.
Expert guidance was used to annotate labels for investigated tail categories. For DCIS labels, we identified 34 DCIS lesions based on pathology results. Additionally, an expert annotated NCM or CAL labels on the 1,387 lesions with ``gold standard" labels. From these annotations, the training set included 741 lesions with NCM (117 benign and 624 malignant) and 251 lesions with CAL (36 benign and 215 malignant).
For validation and selected hyper-parameters, we split the training set into five parts to perform 5-fold cross-validation. 
In the internal test set, we retained 579 lesions (274 benign and 305 malignant) with ``gold standard" labels, consisting of 389,066 frames (179,640 benign and 209,426 malignant). To accelerate evaluation, we sparsely sampled 16,076 frames (7,560 benign and 8,516 malignant), ensuring that the time interval between each pair of sampled frames was at least one second (30 frames). This was feasible because we found that lesion-level results remained consistent with using all frames (difference smaller than 0.01\%).

For external evaluation, we prospectively collected 227 lesions (including 7 DCIS lesions) from 225 consecutive patients who underwent breast-US examinations between October 2022 and March 2023 at PUMCH. These 227 lesions were recruited by a group of radiologists and comprised 63 benign and 164 malignant lesions, all with biopsy-confirmed ``gold standard" labels. Since the 7 DCIS lesions were insufficient to evaluate the model's diagnostic performance for DCIS, we purposely collected an additional 133 DCIS cases. These additional DCIS lesions were sourced from two external institutions: 114 from CICAMS, an institution focused on cancer treatment, and 19 from PUMCH, a comprehensive medical institution. The breast-US examinations for these DCIS cases were conducted between January 2022 and April 2023.

\subsection{Development of TAILOR-Gen}
\label{subsec:gen_model}

Here, we introduce the training and sampling process of TAILOR-Gen, as well as the data cleaning process for high-quality generated images. We design TAILOR-Gen as a conditional Denoising Diffusion Probabilistic Model (DDPM)~\cite{ho2020denoising,dhariwal2021diffusion}. To clarify its design, we first explain the mechanism of DDPM. The training process of DDPM enables it to learn the data distribution $P(x)$ of breast-US images. Specifically, DDPM learns to gradually denoise a Gaussian noise sample $x_T\sim \mathcal{N}(0, I)$ to produce an image $x_0\sim P(x)$. This is achieved via learning the reverse process $P(x_{t-1}|x_{t})$ of a Markov Chain of length $T$. DDPM can be interpreted as a denoising autoencoder $\epsilon_{\theta}(x_t,t)$, which estimates the noise $\epsilon$ in $x_t$ at each step. The learning objective is simplified to:
\begin{align}
    \mathcal{L}=E_{x,\epsilon,t}\left[||\epsilon - \epsilon_{\theta}(x_t,t)||_2^2\right]
\end{align}
where $t$ is uniformly sampled from $\{1,\cdots,T\}$. 
TAILOR-Gen uses a conditional DDPM~\cite{ho2021classifierfree} $\epsilon_{\theta}(x_t,t,c)$ to learn the conditional data distribution $P(x|c)$. Here, $c$ can be any conditioning input, such as specific pathology labels, tail categories, device types, and lesion boxes. 

In the fine-tuning step of TAILOR-Gen with limited domain-specific data, we employ several strategies to enhance the quality of the generated data. To preserve the domain knowledge acquired during the pre-training step, we freeze the pre-trained parameters and fine-tune only the additional lightweight parameters designed as low-rank adapters (LoRA)~\cite{hu2022lora}. To prevent overfitting, we apply strong conventional data augmentations, such as random crop, color jittering, and random flip. Additionally, we incorporate a novel device type data augmentation, transforming each image to all device types~\cite{gatys2016image}. This image-to-image translation task is performed using CycleGAN~\cite{zhu2017unpaired} models, trained on every pair of device types.

In the sampling process, we employ classifier-free guidance~\cite{ho2021classifierfree} to achieve better control of the generated images with input conditions. This method can be formulated as:
\begin{align}
    \tilde{\epsilon}_{\theta}(x_t,t,c)=(1 + w)\epsilon_{\theta}(x_t,t,c)-w \epsilon_{\theta}(x_t,t)
\end{align} 
where $\tilde{\epsilon}_{\theta}(x_t,t,c)$ is defined as the weighted combination of conditional and unconditional DDPM outputs, and $w$ is a parameter that controls the strength of the guidance. The generation process of TAILOR-Gen is inherently slow due to the requirement of $T$ denoising steps. To accelerate the generation while maintaining high-quality outputs, we utilize a sampling technique called DPM-Solver~\cite{lu2022dpm}. We ensure a well-balanced and diverse set of conditioning inputs for sampling. The condition selection strategy is detailed in \spsection{2.2}.

Occasionally, TAILOR-Gen can generate low-quality data, because the distribution learned from thousands of lesions is not perfectly aligned with the real data distribution $P(x)$. To address this, we implement a data cleaning process to remove the low-quality generated data. Specifically, we focus on generated images with incorrect pathology labels (i.e., generated lesions that are inconsistent with the given conditions), as these can be particularly detrimental to the training of TAILOR-Diag. As detailed in \spsection{2.3}, we automatically identify images likely to have incorrect pathology labels using data-driven filters. After the data cleaning process, approximately 10\% of the generated data are removed. We observe a notable improvement in the performance of TAILOR-Diag following this data-cleaning step.

\subsection{Development of TAILOR-Diag}
\label{subsec:diag_model}

We design TAILOR-Diag as an ensemble of four classification models to accurately diagnose various cases using specialized knowledge. Let $\{x_i|i=1, \cdots, N\}$ denote $N$ breast-US images of a lesion from $N$ different scanning views. For an input image $x_i$, we first feed it into the general model, TAILOR-Diag-Base, and get the predicted logit $\hat{y}^{\text{base}}_i$ for common cases. Then, three expert models provide their confidence scores $\hat{c}^{\text{k}}_i$ to determine whether they should be used to diagnose $x_i$ where $\hat{c}^{\text{k}}_i\in[0,1]$ and $\text{k}\in\{\text{ncm},\text{cal},\text{dcis}\}$ for TAILOR-Diag-NCM, -CAL, and -DCIS, respectively. We define the confidence scores as the predicted probability of $x_i$ belonging to each tail category and use thresholds $t^{\text{k}}$ to determine whether to use each expert model. The predicted logits from the expert models are denoted as $\hat{y}^{\text{k}}_i$. Subsequently, we aggregate the predictions of the general and selected expert model(s) to obtain the logit $\hat{y}_i$ for image $x_i$:
\begin{align}
    \hat{y}_i = \hat{y}^{\text{base}}_i + \sum_{\text{k}\in \Omega_i} w^{\text{k}}\cdot \hat{y}^{\text{k}}_i
\end{align}
where the selected indices are $\Omega_i=\{\text{k}| \hat{c}^{\text{k}}_i>t^{\text{k}}, \text{k}\in\{\text{ncm},\text{cal},\text{dcis}\}\}$, and the aggregation weights $w^{\text{k}}$ are determined by 5-fold cross-validation. Finally, we aggregate the logits of all $N$ images to obtain the final prediction of the lesion:
\begin{align}
    \hat{p}=\sigma(\frac{1}{N}\sum_{i=1,\cdots, N}\hat{y}_i)
\end{align}
where $\sigma(\cdot)$ is the sigmoid function, and $\hat{p}\in[0,1]$ is the predicted probability of malignancy of the lesion.

\subsection{Hyperparameters}
\label{subsec:hyperparam}

Hyperparameters of TAILOR-Gen and TAILOR-Diag are carefully selected using 5-fold cross-validation on the training set. We train TAILOR-Gen for $70$ epochs on the entire training set and fine-tune TAILOR-Gen for $70$ epochs on the domain-specific data. We use a batch size of $8$ for training and $128$ for sampling. For optimization, we use an AdamW optimizer with an initial learning rate (LR) $6.25\times 10^{-6}$ and weight decay $1.0\times 10^{-4}$. A Cosine Annealing scheduler is applied to decrease the LR progressively. A clipping of gradient value with a threshold of $1.0$ is employed for training stability. In the data generation process, the classifier-free guidance strength $w=1.8$ and the generated image size is $160\times 160$. We set the steps $T=500$ to train the DDPM, and we utilize the DPM-Solver~\cite{lu2022dpm} to speed up sampling with inference steps $T=50$. 

For TAILOR-Diag, we implement the diagnostic model using the largest Swin Transformer (Swin-L). To satisfy the input size requirement of Swin-L~\cite{liu2021swin}, we resize generated images to $224\times 224$. We train the TAILOR-Diag-Base for $5$ epochs and fine-tune expert models for $2$ epochs. We use a batch size of $128$ during training. For optimization, we use an AdamW optimizer with an initial LR of $5.0\times 10^{-5}$ for training TAILOR-Diag-Base, and the fine-tuning LR of $5.0\times 10^{-6}$. A MultiStep scheduler is used to decrease the LR with a multiplier of $0.1$. We set the weight decay to $0.1$ for training TAILOR-Diag-Base and $0.2$ for fine-tuning to prevent overfiting. During training TAILOR-Diag, we apply the data augmentations including random cropping (ensuring complete lesion areas), random horizontal flipping with probability 0.5, and color jittering for brightness and contrast by a randomly chosen factor from $[0.7, 1.3]$. During the evaluation, we set the thresholds for expert model selection to $t^{\text{ncm}}=t^{\text{cal}}=t^{\text{dcis}}=0.9$; and we set the aggregation weights to $w^{\text{ncm}}=w^{\text{cal}}=2.0$ and $w^{\text{dcis}}=1.0$. 

\subsection{Statistical analysis}

We estimate the 95\% confidence intervals by 1,000 bootstrap replications. We calculate the two-sided P-values for significance comparisons of sensitivity and specificity using permutation tests with 10,000 permutations. The P-values of AUC are calculated using DeLong's test~\cite{delong1988comparing,sun2014fast}. 

\subsection{Implementation details}

We implemented the project based on the following packages: Python (3.9), OpenCV (4.9.0.80), Pandas (2.2.1), Numpy (1.26.4), and Pillow (10.3.0). Additionally, the deep learning model is implemented using PyTorch (1.10.1) and Torchvision (0.11.2). Evaluation metrics are calculated using Sklearn (1.4.1). We conduct the experiments using computational resources from 7 GPU clusters. Four of these clusters each consist of 8 NVIDIA RTX 3090 GPUs, while the remaining three clusters each comprise 8 NVIDIA RTX 4090 GPUs.

\backmatter

\bmhead{Data Availability}

Due to respective Institutional Review Boards' restrictions and to protect patient privacy, the training and test datasets used in this study cannot be made publicly available. The BUSI test dataset used in this study is publicly available at \url{https://scholar.cu.edu.eg/?q=afahmy/pages/dataset}.

\bmhead{Acknowledgments}
We thank Ruichen Li and Jigang Fan for their helpful suggestion and discussion.
Liwei Wang is supported by the National Science and Technology Major Project (2022ZD0114902) and National Science Foundation of China (NSFC62276005). Di He is supported by National Science Foundation of China (NSFC62376007).

\bmhead{Author contributions}

H.Y. and L.W. conceived and designed the study. 
Z.N., B.T., Y.Lu. and X.G. carried out data acquisition. 
Y.Li., H.Y., Y.Lu., Z.N. and Q.W. carried out data processing and annotation.
H.Y. developed the AI models.
Y.Li. carried out generated data cleaning.
Y.Li. developed the platform for reader study.
N.Z., Z.N., W.Q., J.T., M.Z., X.G., J.H., L.H. and Y.W. participated in the reader study.
H.Y., D.H., Y.Li., N.Z., Z.N., D.W., Z.Z., Q.W., D.D., Q.Z. and L.W. wrote and revised the paper.

\bmhead{Competing Interests}

The authors declare no competing interests.
\bibliography{sn-bibliography}
\end{document}